\documentclass{aa}
\usepackage{graphicx}

\begin{document}
%

   \title{Is the cluster of galaxies ``Abell 194'' surrounded by an Einstein-Straus vacuole?}

   \author{R. Plaga\inst{1}}

   \offprints{Rainer Plaga}

   \institute{Franzstr. 40, 53175 Bonn, Germany\\
              \email{rainer.plaga@gmx.de}}

   \date{Received ; accepted }

   \authorrunning{R. Plaga}
   \titlerunning{A194 surrounded by an Einstein-Straus vacuole?}

   \abstract{It is suggested that
an ``Einstein-Straus vacuole'' -- a region of
space time with a metric obtained by solving
the equations of general relativity of a mass condensation in an expanding
universe with vanishing cosmological constant -- 
surrounds the cluster of galaxies ``Abell 194''.
This hypothesis is shown to predict a distribution
of galaxy redshifts that is in better accord with
observations than the one expected in the
cosmological concordance model.

      }

   \maketitle
%

60 years ago Einstein and Straus (1945,1946) determined the metric of
space-time near a star embedded in an expanding universe without a cosmological constant
within general relativity (GR).
They assumed the star with mass M is surrounded by an empty spherical cavity with
a radius: 
\begin{equation}
r_{\rm ES} =\left({3 M}\over{4 \pi \rho_{\rm M}}\right)^{1/3}.
\label{es}
\end{equation}
Here $\rho_{\rm M}$ is the mean matter density of the universe. The initially homogeneous matter within
the cavity can be thought to be ``condensed into the star''.
\\
Einstein and Straus found that space-time in the interior of the cavity is described by a static
Schwarzschild metric.
The cavity boundary expands with the unimpeded Hubble flow. 
Outside the cavity space-time is described by the expanding
Friedmann-Lemaitre metric of a homogeneous universe.
The ``static blop of vacuum with Schwarzschild metric'' that surrounds the 
star has been called ``Einstein-Straus vacuole''
(Bonnor 1987). 
With a non-vanishing cosmological constant $\Lambda$ 
eq. (\ref{es}) remains valid, but the metric 
within this ``$\Lambda$ vacuole'' is
the Schwarzschild-de Sitter metric
(for a summary
see Gilbert (1956)). In its Newtonian limit one finds that a new force directed away
from the central mass should exceed gravitational attraction beyond
a radius
\begin{equation}
r_{\rm m} = \left({3 G M \over {\Lambda}}\right)^{1\over 3}.
\label{eq-m}
\end{equation}
The distance from a central mass that is embedded in an expanding universe
at which test particles just begin to follow the outward
Hubble flow rather than being attracted towards the centre
is commonly called the ``turnaround radius'' $r_{ta}$.
In an Einstein-Straus vacuole this transition takes place at the
outer boundary of an Einstein-Straus vacuole, symbolically
\begin{equation}
r_{\rm ta}({\rm Einstein-Straus \ vacuole})=r_{ES},
\label{eq-taES}
\end{equation}
whereas for the $\Lambda$-vacuole outward expansion sets in
at $r_m$ and thus:
\begin{equation}
r_{\rm ta}({\rm \Lambda \ vacuole})=r_{\rm m}.
\label{eq-tam}
\end{equation}
For the cosmological concordance model (Spergel et al.,2003) (
for which the parameters H$_0$=70 km/sec/Mpc and 
$\Omega_{\rm M}$=0.3, $\Omega_{\Lambda}$=0.7 are assumed throughout
this paper) $r_{\rm ES}$ exceeds $r_{\rm m}$ by about
a factor 1.7 at the present epoch.
Vacuoles
do not surround stars like our sun, because they are already embedded in a static space time
within our Galaxy. Clusters of galaxies, among the largest known structures that are gravitationally
bound, could be surrounded by vacuoles.
Within the concordance model an isolated fully collapsed cluster is expected to be surrounded
by a $\Lambda$ vacuole. Clusters might not
have completed their collapse at the present time.
This would reduce
the turnaround radius as compared the value predicted in eq.(\ref{eq-tam}) and one
concludes:
\begin{equation}
r_{\rm ta}({\rm galaxy \ cluster, today}) \leq r_{\rm m}.
\label{eq-itam}
\end{equation}
The medium-compact cluster Abell 194 (A194) at a distance of about 76 Mpc  is fairly 
isolated and its central galaxies have approached energy equipartition.
Its central region is characterised by an elongated line of bright galaxies, but its outer
region is spherical (Zwicky $\&$ Humason, 1964).   
The redshifts of 6916 galaxies around this cluster have been collected and
analyzed in the recent CAIRNS survey (Rines et al. 2003).
In a scatter plot against clustercentric radius r they display a distinct density enhancement around
the cluster redshift z=0.0178 within a well defined boundary. The redshift amplitude of the
boundary is observed to decrease
with clustercentric radius $r$ from $\Delta$z = z(boundary) - 0.0178 $\approx$ $\pm$ 4 $\times$ 10$^{-3}$ 
at $r$=0
to zero at a distance of $r_{\rm c}$ $\approx$ 13.2 Mpc. 
This trumpet shaped ``caustic'' boundary (with a ``mouth piece'' at $r_c$)
had been theoretically predicted for galaxies that fall radially
onto a cluster (Kaiser 1987). 
The pattern is smoothed out, but does not vanish, if galaxies bound to the cluster
possess random motions perpendicular to the infall direction (Diaferio 1999).
In theoretical calculations the dispersion of redshifts becomes minimal
at the turnaround radius $r_{\rm ta}$.
This is because at this radius - excluding random motions -
infalling or orbiting galaxies are at rest relative to the 
cluster and therefore have the mean cluster redshift, i.e. $\Delta$z=0.
Therefore - if observationally the caustic boundary is a distinct, continuous structure -
it is believed, that the galaxies do not follow the Hubble flow at clustercentric distances
below $r_{\rm c}$ (Diaferio 2005).
Symbolically the observations of A194 seem to indicate:
\begin{equation}
r_{\rm c}(A194) \approx 13.2 \ {\rm Mpc} \leq r_{\rm ta}(A194).
\label{caustic}
\end{equation}
For an observed total cluster mass 
\begin{equation}
M(A194) = (2.9 \pm 2.2) \times 10^{14} M_{\odot}
\label{mass}
\end{equation}
(Rines et al., 2003) one finds by inserting $\Lambda$=8$\pi$G$\rho_{\Lambda}$ into 
eq.(\ref{eq-m}) and restating eq.(\ref{eq-itam}):
\begin{equation}
r_{\rm m}(A194) =
6.9_{-2.6}^{+1.4} \ {\rm Mpc} \geq r_{\rm ta}(A194).
\label{attr}
\end{equation}
This inequation disagrees with the one in eq.(\ref{caustic}) by more
than four times the stated error. The observed
turnaround radius of A194 is much larger than the upper limit from eq.(\ref{attr}) that
would seem to be a firm prediction of the concordance model of cosmology.
\\
One possible explanation for this discrepancy
is that the observed caustic boundary around A194
is due to galaxies that are related but unbound to the cluster
and which coincidentally have redshifts consistent 
with being in an infall pattern onto
A194, i.e. the observed shape of
the caustic pattern is
due to chance at least beyond $r_{\rm m}$ (Rines 2005).
\\
An alternative
hypothesis, called ``ES'', is:
\\
``{\it Abell 194 is surrounded by an Einstein-Straus vacuole with a radius
of about 13.2 Mpc. The mass within the vacuole has nearly completely collapsed
onto the central virialized cluster. The vacuole contains a static Hernquist
halo with galaxies on isotropic orbits.}''
\\
Assuming $\Omega$=0.3, eqs. (\ref{es},\ref{mass}) yield:
\begin{equation}
r_{\rm ES}(A194) = 11.7_{-4.4}^{+2.4} \ {\rm Mpc}. 
\label{es_a194}
\end{equation}
If hypothesis ES is true it is
plausible that the observed $r_{\rm c}$ 
corresponds to the turnaround radius and
eq.(\ref{eq-taES}) predicts:
\begin{equation}
r_{\rm ta} = r_{\rm ES} \approx r_{\rm c}. 
\label{eq-hyp}
 \end{equation}
This 
relation is fulfilled by the 
numerical values of eqs.(\ref{caustic},\ref{es_a194}) within the
stated 1-sigma errors.
Rines et al.(2003) showed that the observed redshifts around A194
could well be due to
a radial mass- and velocity-dispersion distribution 
expected in a static Hernquist halo.
The mean halo density of A194 near $r_{\rm c}$ 
in the Hernquist model is about 60 times smaller
than the mean density of the universe, i.e. the outer halo 
would be a good approximation to the ``empty cavity'' within
the Einstein-Straus vacuole.
\\
Hypothesis ES thus seems to describe observations
well and is a solution of the GR field equations in a $\Lambda$=0 universe.
\\
What would be the implications for the nature of $\Lambda$ in general
if hypothesis ES were true?
The simplest (but not only\footnote{
In principle e.g. modifications
of GR could make hypothesis ES true, even with a concordance $\Lambda$.}) modification of the concordance
model that would render possible hypothesis ES would be
that the cosmological constant vanishes or at least
does not act on distance scales up to about 20 Mpc.
While the observational evidence for an accelerated cosmic expansion
seems compelling, its physical origin is still a matter of debate.
Besides a cosmological constant some negative-pressure fluid with
adjustable properties (``dark energy'') might be responsible.
Recently Kolb et al.(2005) even proposed an acceleration mechanism that
does not rely on a cosmological constant or
any other ``dark energy'' component.
Therefore
a modified $\Lambda$ that does not act at least up to cluster distance scales does not seem to 
be in contradiction with
firmly established facts. However, the case for such a far reaching
conclusion from the observation of just one cluster of galaxies
remains weak.
\\
Can Einstein-Straus vacuoles form around clusters of galaxies if $\Lambda$=0?
Even with a vanishing $\Lambda$ the 
idealized spherical-infall model of cluster formation
predicts a turnaround radius for A194 at the present time 
of $r_{\rm ta}$(spherical infall, A194) $\approx$ 4.8 Mpc (Rines et al., 2003), 
which is far smaller than $r_{\rm ES}$(A194).
However,
more realistic simulations (e.g. (Diaferio, 1999)) show that cluster formation
occurs very anisotropically and involves effects
like shell crossings not taken into account in the spherical-infall model.
Therefore $r_{\rm ta}$(spherical infall) should only be
considered as an indicative but not quantitative estimate (Diaferio 2005).
Other modifications of the concordance model than a vanishing $\Lambda$
would have an unknown influence on the collapse process.
If some fraction of clusters of galaxies is expected to
have completed their collapse by now -- as required by hypothesis ES --
remains unclear presently.
\\
Further observations
of complete caustics around relaxed, isolated clusters 
of galaxies could test hypothesis ES.
E.g. the Coma cluster 
observationally displays definite caustic boundaries (Rines et al., 2003)
up to a clustercentric radius of at least up to about 14 Mpc.
If the Coma cluster is surrounded by an Einstein-Straus vacuole
$r_{ta}$(Coma) $\approx$ 20 Mpc is predicted\footnote{For a cluster mass of 
2 $\times$ 10$^{15}$ M$_{\odot}$},
whereas the concordance model (eq.(\ref{eq-itam})) predicts that $r_{\rm ta}$(Coma) $\leq$ 13 Mpc.
\\
Should further observations confirm the existence of
Einstein-Straus vacuoles around a fraction of 
clusters of galaxies, this would indicate that at least one fundamental 
theoretical element is still missing in the present concordance model of
cosmology. The most conservative (but not only) options would seem to be
a vanishing $\Lambda$ or an appropriately modified form of dark energy.

\begin{acknowledgements}
I thank Antonaldo Diaferio and Kenneth Rines 
for very helpful correspondence on the subject of this paper
and reading a previous version of the manuscript.
The constructive suggestions of an anonymous referee 
greatly improved the manuscript. Stacy McGaugh and Moti
Milgrom patiently answered questions 
about the structure of spacetime in an expanding universe.
     
\end{acknowledgements}

\end{document}